\documentstyle[11pt,newpasp,twoside, psfig]{article}
\newcommand{\Msun}{\ifmmode {M_{\odot}}\else${M_{\odot}}$\fi~}
\newcommand{\Rsun}{\ifmmode {R_{\odot}}\else${R_{\odot}}$\fi}

\def\mdot{$\dot{m}$~}  

\def\edcomment#1{\iffalse\marginpar{\raggedright\sl#1\/}\else\relax\fi}
\reversemarginpar

\begin{document}
\title{Variability and Spectra of Two Neutron Stars in 47 Tucanae}
 \author{C. O. Heinke, J. E. Grindlay, D. A. Lloyd, P. D. Edmonds}
\affil{Harvard-Smithsonian Center for Astrophysics, 60 Garden Street, Cambridge, MA  02138}

\begin{abstract}
We report spectral and variability analysis of two 
quiescent low mass x-ray binaries (previously identified with ROSAT
HRI as X5 and X7) in the globular cluster 47 Tuc, from a 
{\it Chandra} ACIS-I observation.   X5 demonstrates sharp eclipses with an
8.666$\pm0.008$ hr period, as well as dips 
showing an increased $N_H$ column.  
Their thermal spectra are
well-modelled by unmagnetized hydrogen atmospheres of hot neutron
stars, most likely heated by transient accretion, with little to no
hard power law component.
\end{abstract}

\index{accretion disks}
\index{binaries: close}
\index{binaries: eclipsing}
\index{binaries: x-ray}
\index{globular clusters: individual (NGC 104)}
\index{stars: neutron}

\section{Introduction}

Many neutron stars (NSs) that have been observed in outburst as soft x-ray
transients have also been detected in quiescence ($L_X\sim10^{32-34}$
ergs s$^{-1}$), for example Cen X-4 and Aquila X-1 (see Campana et
al. (1998a) for a review of soft x-ray transients,
also known as quiescent low-mass x-ray binaries (qLMXBs). Spectral
fits with a soft 
(kT=0.2-0.3 keV) blackbody (BB) spectrum, often 
requiring a hard power-law tail of photon index $\sim2$, have been
acceptable but imply an emission area of $\sim1$ km radius, smaller
than a NS surface.  

However,
Rajagopal \& Romani (1996) and Zavlin et al. (1996) showed that the
atmosphere of a NS, due to the strong frequency dependence of
free-free absorption, shifts the peak of the emitted radiation to
higher frequencies; thus a blackbody fit will derive a temperature
that is too high and a radius that is too small.  Brown, Bildsten,
and Rutledge (1998; BBR) showed that the interior of a 
transiently accreting neutron star is heated during accretion by
pycnonuclear reactions, bringing the interior to a steady state
temperature $\sim10^8 \langle$ \mdot$/10^{-10} M_{\sun}$
yr$^{-1}\rangle^{0.4}$ K. This heating leads to an isotropic  
thermal luminosity between 
accretion episodes of roughly $L_q=6\times10^{33}$ ergs s$^{-1}
(\langle$ \mdot $\rangle/10^{-10}$ \Msun yr$^{-1}$ (BBR).
Fits of simulated low magnetic-field ($B<10^{10}$ G) hydrogen
atmospheres to qLMXBs have given implied 
$R_{\infty}$ of roughly 13 km ($R_{\infty}$ is the effective radius
seen by a distant observer, $R_{\infty}=R/\sqrt{1-2GM/Rc^2}$), with a
power law component in Cen X-4 and Aql X-1 suggested to be
linked to continued accretion (Rutledge et
al. 2001a, 2001b, Campana et al. 1998b).

The deep ROSAT HRI observation of 47 Tucanae of Verbunt \& Hasinger (1998)
resolved 5 sources, including X5 and X7, in the central core, but
without spectral resolution.  With {\it Chandra}'s
ACIS-I instrument, we have identified 108 sources in the central
2\arcmin$\times$2.5\arcmin region, and are able to conduct spectral
analysis on a few dozen of these sources (Grindlay et al. 2001a).  X5
has been identified  with a $V=21.6$ counterpart,  
which shows variability and a blue color indicative of a faint accretion
disk, while a tentative upper limit for the 
counterpart of X7 is $V\sim23$ (Edmonds et al. 2001).  Using
$F_{V}=10^{-0.4V-5.43}$ (using the 5000-6000 \AA range) gives $F_X/F_V$ ratios
of 43 and $>$166 
respectively, which are typical of LMXB systems rather than CVs
(typically $\sim1$).  The
likely qLMXBs X5 
and X7 have blackbody-like x-ray spectra (see below) which show they are indeed
hot NSs.

\section{Variability}

The {\it Chandra} X-ray Observatory observed 47 Tucanae on March
16-17, 2000, for 72 ksec (in five contiguous exposures)
with the ACIS-I instrument at the focus.  
We derive luminosities for X5 (including eclipses and dips) and
X7 of $L_X$(0.5-2.5 
keV)=$7.4\times10^{32}$ ergs s$^{-1}$, $9.0\times10^{32}$ ergs 
s$^{-1}$ respectively, consistent with the ROSAT observations.

\begin{figure}

\psfig{file=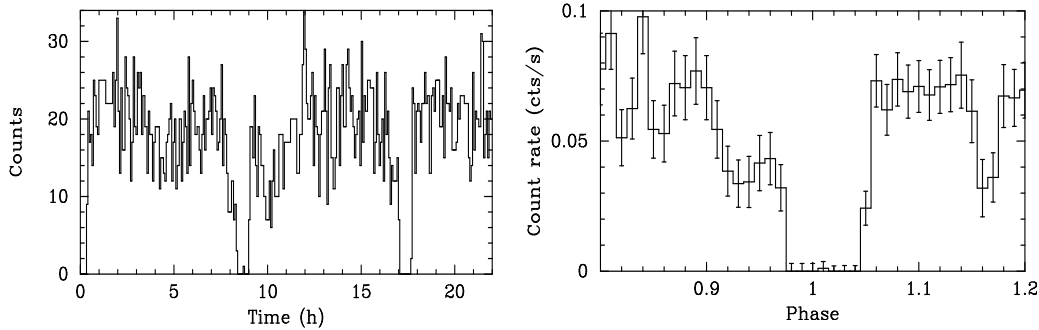}

\vspace{0.5cm}

\caption[heinkec2_1.eps]{Top: Lightcurve of X5, showing the three
eclipses and other dips. Bins are each 264
seconds long.  Bottom: Phase plot
of the eclipse 
and pre-eclipse dip portion of X5's lightcurve, using 8.666 hour
period. Bins are 156 s long.\label{Figure 1}} 
\end{figure}

The
lightcurve of X5 shows significant variability (Figure 1), in particular 
three clear eclipses and significant dips (99\% significance for
 variability outside eclipses, according to a K-S test).
X7 shows no significant variability.
The three egresses of X5's eclipses are quite 
sharp (less than 20 seconds, see Fig. 1a), whereas the two ingresses
show a dip of roughly 
half the normal flux for $\sim2000$ seconds before the eclipse.
Using the 3.2 s ACIS time resolution for the sharp egresses, a value 
of 8.666$\pm$.008 hours may be estimated for the
period.  The eclipse lengths for the second two eclipses are
2482$\pm$30 seconds.  The dips
directly before the eclipses suggest an accretion stream obscuring the
NS, while the other dips and concomitant 
increases in the $N_H$ column are typical of occulting blobs in the
edge of an accretion disk, as seen in dip sources. Assuming a 1.4
$M_{\sun}$ NS and a 0.3 $M_{\sun}$ secondary (to which the calculation
is relatively insensitive), the separation is
$1.76\times10^{11}$ cm, and the 
secondary must have a diameter of at least $8.8\times10^{10}$ cm to
perform the eclipse.  The restriction on not filling its Roche lobe,
plus the eclipse duration, give a minimum secondary mass of 0.27
$M_{\sun}$, for a 90 degree inclination and 1.4 $M_{\sun}$ primary,
regardless of the interior structure of the secondary.  The size of
X5's secondary rules out a helium white dwarf, and thus indicates that the
secondary is likely to be a main-sequence star.

\section{Spectra}

We fit the spectra of X5 and X7 in XSPEC, using the 
unmagnetized hydrogen-atmosphere models of Lloyd, Hernquist, \& Heyl (2001,
and in these proceedings).  The
atmospheric plasma was assumed to be completely ionized, and the
spectrum due to coherent electron scattering and free-free absorption. 
Initial fits were made using the surface gravity of a canonical NS,
$\log{g_{s}}=14.38$ (appropriate for a 1.4 $M_{\sun}$, 10 km NS).
Magnetic fields up to $10^{11}$ G will 
not affect these fits in either flux or $T_{eff}$ (Lloyd et al.,
2001).  We use the pileup formalism of J. Davis (2001)
incorporated into XSPEC, since X5 and X7 suffer 9 and 10\% pileup,
respectively. Photoelectric absorption was a free parameter, using the
47 Tuc metallicity, and a residual feature was modeled with an edge
around 0.63 keV (perhaps due to hot gas), reducing $\chi^2_{\nu}$ by 0.3.

\begin{figure}
\psfig{file=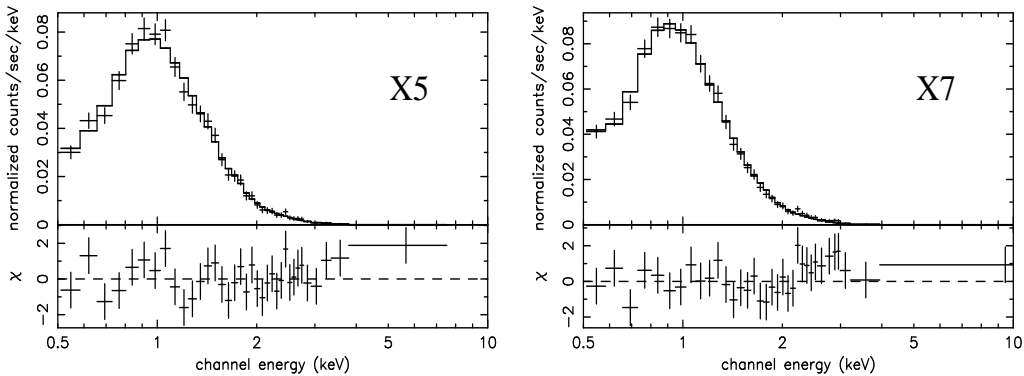}

\vspace*{0.5cm}

\caption[heinkec2_2.eps]{Data and model hydrogen-atmosphere spectra (with
Davis pileup model) of X5 (left) and X7 (right), between 0.5 and 10 keV.
\label{Figure 2}}
\end{figure}

The spectra of X5 and
X7 along with the H-atmosphere model predictions and the fit residuals
are plotted in Figure 2.  Both qLMXBs give good fits, 
$\chi^2_{\nu}=0.98$ for 30 degrees of freedom.  Unfortunately we
cannot meaningfully constrain the radius alone, but do constrain a
region in M-R space; details of the spectral fitting and implications 
for the masses of X5 and X7 are given in Heinke et
al. (2002, in preparation). For the range of radii generally assumed
for neutron stars (9-14 km), we derive temperatures between 92 and 108
eV, and $N_H=2.5^{+0.9}_{-0.5}\times10^{21}$ cm$^{-2}$. 

The qLMXBs Cen X-4 and Aql X-1 have generally shown a hard power-law
 tail, of photon index 1-1.5, dominant above 2 keV.  Such a tail is nearly
 absent from X5 and X7, where a PL of index 1.5 is constrained to supply less
 than 5\% of the flux between 0.5 and 10 keV, in either fit. A similar lack of
 observable PL tail has been seen in the qLMXB CXOU 132619.7-472910.8
 in $\omega$ Cen (Rutledge et al. 2001c), and in NGC 6397 (Grindlay et
 al. 2001b).

We interpret the lack of a power law
and variability
 as signs that accretion onto the NS surface is held at a very low
level or completely stopped.  Yet accretion to the disk is clearly
continuing in X5, as shown by the dips and blue color of the companion. 
This paradox may be solved by the thermal/viscous instability if
little to no ADAF flow is currently taking place (Dubus et al. 2001
and refs therein).  Wijnands et
al. (2001) suggest that to explain the low luminosity of KS 1731-260
in quiescence after a decade of active accretion, it must spend hundreds of
years in quiescence between 
outbursts, and postulate a population of qLMXBs with long recurrence
times and long outbursts.  Perhaps these systems and KS
1731-269 have extremely low disk viscosity, and thus long recurrence
times and long outbursts due to greater disk accumulation.  
Upcoming longer observations
with Chandra will further constrain the spectra of
these objects. 

\acknowledgments

This work was supported in part by Chandra grant GO0-1034A.
C.H. thanks P. Kondratko, J. McDowell, J. Houck, H. Marshall,
J. Raymond, and J. Davis.

\end{document}